\documentclass[aps,nofootinbib,prd]{revtex4} 
\usepackage{amsfonts}
\usepackage{amssymb}
\usepackage{graphicx}
\usepackage{pstricks}
\usepackage{epsfig}

\newcommand{\usual}{1-(n-m+1) \frac{q}{3D} \frac{\partial D}{\partial
q}} \newcommand{\be}{\begin{equation}}
\newcommand{\ee}{\end{equation}}

\begin{document}

\title{On the Onset of Inflation in Loop Quantum Cosmology}

\author{Cristiano Germani}

\affiliation{SISSA and INFN, via Beirut 4, 34014 Trieste, Italy}
%\ead{Germani@sissa.it}

\author{William Nelson}

\affiliation{Department of Physics, King's College London, London WC2R
2LS, England} 
%\ead{William.Nelson@kcl.ac.uk}

\author{Mairi Sakellariadou}

\affiliation{Department of Physics, King's College London, London WC2R
2LS, England} 
%\ead{Mairi.Sakellariadou@kcl.ac.uk}

\begin{abstract}
Using a Liouville measure, similar to the one proposed recently by
Gibbons and Turok, we investigate the probability that single-field
inflation with a polynomial potential can last long enough to solve
the shortcomings of the standard hot big bang model, within the
semi-classical regime of loop quantum cosmology. We conclude that, for
such a class of inflationary models and for natural values of the loop
quantum cosmology parameters, a successful inflationary scenario is
highly improbable.
\end{abstract}

\maketitle
\section{Introduction}

Cosmological inflation~\cite{infl1} is at present the most promising
candidate to solve the shortcomings of the standard hot big bang
model, although other mechanisms have been
proposed~\cite{stringgas,cyclic,sling}. Inflation essentially consists
of a phase of accelerated expansion which took place at a very high
energy scale.  One of the appealing features of inflation is that it
is deeply rooted in the basic principles of general relativity and
field theory. In addition, when the principles of quantum mechanics
are taken into account, inflation provides a simple explanation for
the origin of the large scale structures and the associated
temperature anisotropies in the cosmic microwave background radiation.

Despite its success, inflation is still a paradigm in search of a
model and its strength is based on the assumption that its onset is
generically independent of the initial conditions.  However, even when
the issue of the onset of inflation was
addressed~\cite{piran,calzettamairi}, no robust conclusions could be
drawn as a quantum theory of gravity was missing.  Recently, it has
been argued~\cite{GT} that the probability of having $N$ (or more)
e-foldings of inflation within single-field, slow-roll inflationary
models is suppressed by an order of $\exp(-3N)$. However, in finding
this result the authors have used a classical theory even at energy
scales for which the quantum effects can no longer be
neglected. Moreover, as we shall discuss, the analysis of
Ref.~\cite{GT} is not always valid.  In what follows, we estimate the
probability to obtain a sufficiently long inflationary era, in the
context where not only general relativistic but also quantum effects
are taken into account. More precisely, we study the probability of
having a sufficiently long inflationary era in the context of loop
quantum cosmology.

We organize the rest of the paper as follows. In Section 2, we briefly
discuss some elements of loop quantum cosmology. In Section 3, we
discuss inflation within the loop quantum cosmology framework. In
Section 4, we study the probability to have successful inflation
within this context. We round up with our conclusions in Section 5.

\section{Loop quantum cosmology}

Loop quantum gravity~\cite{rovelli,tie} is at present the most
developed approach to a background independent and non-perturbative
quantization of general relativity, which can deal with the extreme
conditions realized at classical singularities.  The full theory is
still not completely understood, and in a number of cases not even the
continuum limit of space-time can be explicitly found.  Nevertheless,
by introducing symmetries, one may resolve the theory at a
non-perturbative level. More precisely, applying loop quantum gravity
to homogeneous and isotropic cosmologies, the theory becomes
analytically tractable and loop quantum cosmology can be
studied~\cite{martin}. It is worth noting that such mini-superspace
models may not encompass all features of the full inhomogeneous
theory, however it is reasonable to expect that they have, at least
qualitatively, the correct behaviour~\cite{ABL}.

In loop quantum cosmology, the evolution of the universe is divided
into three distinct phases~\cite{martinetal}, depending on the value
of the scales probed by the universe. At first, very close to the
Planck scale, the concept of space-time has no meaning, full quantum
gravity is the correct framework and the universe is in a discrete
quantum phase. Applying loop quantum cosmology during this phase, one
gets a finite bounded spectrum for eigenvalues of inverse powers of
the three-volume density, which we shall call shortly ``the geometrical
density''.  As the volume of the universe increases with time, the
universe enters a semi-classical phase.

For length scales above $L_{\rm Pl}\equiv \sqrt\gamma l_{\rm Pl}$
($\gamma\approx .2735$ is the Barbero-Immirzi parameter and $l_{\rm
Pl}$ denotes the Planck length, with $l_{\rm Pl}^2=8\pi G$
\footnote{Units $\hbar=c=1$ are used in this paper.}), the space-time
can be approximated by a continuous manifold and the equations of
motion take a continuous form, which differs from the classical
behavior due to the non-perturbative quantization effects.  This
intermediate phase is characterized by a second length scale
$L_\star$, with $L_\star\equiv \sqrt{(\gamma j \mu_0)/3} l_{\rm Pl}$,
which determines the size below which the geometrical density is
significantly different from its classical form.  For length
scales below $L_\star$, quantum corrections can no longer be
neglected. The half-integer $j$ labels the ambiguity in choosing the
representation in which the matter part of the Hamiltonian constraint
for a scalar field is quantized. The length parameter $\mu_{0}$,
related to the underlying discrete structure, is the scale of the
finite fiducial cell that spatial integration is restricted to, so as
to remove the divergences that occur in non-compact
topologies~\cite{Vandersloot}. As it was shown in Ref.~\cite{ABL}, one
can use an arbitrary value of $\mu_{0}$. However, one should keep in
mind, that the same value should be adopted for both the Hamiltonian
constraint and the inverse volume operator.  In what follows we set,
for simplicity, $\mu_{0}=1$; as $\mu_0\sim {\cal O}(1)$, different
values of $\mu_0$ do not sensibly modify our conclusions. For $j>3$,
the two scales $L_{\rm Pl}$ and $L_\star$ overlap and space-time can
be considered as continuous. 
The intermediate phase is the most important one regarding the
phenomenological consequences of loop quantum cosmology as it may lead
to distinct signatures~\cite{Tsujikawa:2003vr}.  At later times, and
therefore larger scales, the universe enters the full classical phase
and standard cosmology becomes valid.  The main feature of loop
quantum cosmology is the resolution of the cosmological singularity.
Indeed, one can show that, upon quantization, the operator associated
with the inverse of the three volume never diverges.

The metric of a Friedman-Lema\^{i}tre-Robertson-Walker (FLRW)
space-time reads
\begin{equation}
\label{metric}
{\rm d}s^2= -{\rm d}t^2+a^2(t)\delta_{ij} {\rm d}x^i {\rm d}
x^j=a^2(\eta) [-{\rm d}\eta^2+\delta_{ij}{\rm d}x^i{\rm d} x^j]~,
\end{equation}
where $t\ (\eta)$ is the cosmological (conformal) time (with ${\rm
d}t=a{\rm d}\eta$), $a$ the scale factor and $\delta_{ij}$ the
Kr\"onecker symbol. The geometric density is therefore $a^{-3}$.
Promoting $a^{-3}$ and its inverse ($a^{3}$) at the level of operators,
we have that, upon quantization\footnote{We define $\hat A$ to be the
operator associated with the function $A$.},
 \be
 \langle \hat
a^3\rangle=a^3\ ,\ \langle \hat a^{-3}\rangle=d_{j,\,l}(a)\ ,
 \ee
where the modified density, $d_{j,\,l}(a)$, in the continuum limit of
loop quantum cosmology is given by the following approximated
expression
\begin{equation}
    d_{j,\,l}(a)=D_l(q)a^{-3}~~\mbox{with}~~q=a^2/a^2_\star~.
\end{equation}
The parameter $l$ determines the
behavior of the effective geometrical density on small scales with
respect to $L_\star$ and $l\in [0,1]$.  However, some values of $l$
are preferred and only a discrete sequence is used, $l_k=1-(2k)^{-1}$
with $k\in N$. The function $D_l(q)$ can be written
as~\cite{martinetal}
\begin{equation}
 \label{eq:MB}
    D_l(q)=\biggl\{\frac{3}{2l}q^{1-l}
\biggl[\frac{(q+1)^{l+2}-|q-1|^{l+2}} {l+2}-
-\frac{q}{1+l}\biggl((q+1)^{l+1}-\mbox{sgn}(q-1)|q-1|^{l+1}\biggr)
\biggr]\biggr\} ^{3/(2-2l)}~.
\end{equation}
\begin{figure}
  \begin{center}
      \includegraphics[scale=0.25]{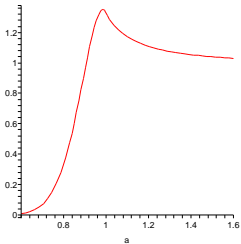}
      \caption{\label{fig3} $D(q)$
is plotted as a function of $a$, with
 $a_{\star}=1$ and $l=0.75$. For $a>>a_{\star}$ $D \rightarrow 1$, giving
us the classical result.}
  \end{center}
\end{figure}
For all allowed values of the parameters $j,l$ the effective
geometrical density behaves as $d_{j,\,l}(a)\rightarrow a^{-3}$, which
is the classical behavior, as $a\gg a_\star$. Around $a_\star$ the
effective geometrical density $d_{j,\,l}(a)$ becomes maximal (see
Figure (\ref{fig3})). For $a\ll a_\star$ the density approaches zero,
\begin{equation}
    d_{j,\,l}(a)\sim \biggl( {3\over 1+l} \biggr)^{{3\over
    2-2l}}\biggl( {a\over a_\star }\biggr)^{{3(2-l)\over 1-l}}
    a^{-3}~,
\end{equation}
resolving the singularity present in the classical theory as $a
\rightarrow 0$ (see Figure \ref{fig1}).  The parameters $j$ and $l$
can only be weakly restricted by considerations of the discrete
structure of the theory. These parameters can in principle be fixed by
knowing the full loop quantum gravity; we will not consider them here
as ambiguities\footnote{These two parameters capture the typical
properties of functions such as $D(q)$, which is the position of the
peak ($j$) and the power law increase at small $q$ ($l$). As we
mentioned previously, other ambiguities in quantitative details exist,
which at the current stage are not relevant for an effective
analysis. Precise functions would follow by relating isotropic models
to the inhomogeneous situation~\cite{aps06,mb06}. We will back shortly
to this issue later.}.  However, as we shall see, there is a more
dangerous ambiguity coming from the fact that $\langle \hat
a^3\rangle\neq 1/\langle \hat a^{-3}\rangle$.

\subsection{Dynamics and ambiguities}

Let us first briefly, and in a rather schematic way, discuss the issue
of quantization ambiguities.  Dynamics are controlled by the
Hamiltonian constraint, which classically gives Friedmann equation.
The Hamiltonian of the whole system, gravity plus matter,
reads~\cite{martinetal}
\begin{equation}
\label{ham}
    {\cal H}\equiv -3\dot a^2 a+8\pi G {\cal H}_{\rm m}\ ,
\end{equation}
where the first term in the {\sl r.h.s} is the gravity part and ${\cal
H}_{\rm m}$ is the matter Hamiltonian. The equations of motion are
satisfied requiring quantum mechanically that $\hat{\cal{H} }|\Psi
\rangle=0$, where $|\Psi\rangle$ is the {\sl wave function} of the
universe and $\hat{\cal{H}}$ is the promotion to an operator of the
classical Hamiltonian.  Semi-classically, this implies $\langle
\Psi|\hat{ {\cal H}}|\Psi\rangle=\langle{\cal H}\rangle=0$. However,
as we shall see, we have ambiguities on deciding which composite
operator of the geometrical density (such as the Hamiltonian) is the
correct one to be used.

In principle, instead of the Hamiltonian, one may consider the
classically equivalent operator
\begin{equation}
\hat{\cal Q}=-3 \hat {a}\hat{\dot a}^2+8 \pi G {\cal H}_{\rm m}
\hat{a}^n \hat{a}^{-n}\ .
\end{equation}
Requiring that classical symmetries, such as diffeomorphism invariance,
are not broken at the quantum level, one obtains that $\hat{\cal Q}$
should, as well as the Hamiltonian, define diffeomorphism invariance,
implying
\begin{equation}
\hat{\cal Q}\ |\Phi\rangle=0\ ;
\end{equation}
$|\Phi\rangle$ is a new state associated with $\hat {\cal Q}$
\cite{martinetal}. In this case however, the semi-classical equations
$\langle \hat {\cal Q}\rangle=0$ differ from $\langle \hat {\cal
H}\rangle=0$ during the quantum regime. Although the difference is at
the quantum level, the trajectories defined by $\langle\hat {\cal
Q}\rangle=0$ and $\langle\hat{\cal H}\rangle=0$ coincide at the
classical limit (as $a\rightarrow\infty$).  Therefore, as the two
evolutions are indistinguishable at the classical level, it is
impossible to decide whether or not the state $|\Psi\rangle$ is more
fundamental than the $|\Phi\rangle$ one.  This is a dangerous
ambiguity in loop quantum cosmology
as it appears whenever, at the quantum level, $\langle \hat
a^3\rangle\neq 1/\langle \hat a^{-3}\rangle$.

An extra ambiguity appears in the choice of the representation in
which the gravity part of the total Hamiltonian is quantized.  This
ambiguity is defined by a similar parameter used to label the
representation for the quantization of the matter Hamiltonian, we will
call this parameter $j_{\rm G}$~\cite{Vandersloot}. As in the matter
part, one can in fact define a new scale $L_{\rm G}=\sqrt{\gamma
j_G\mu_0/3}l_{\rm Pl}$ above which quantum corrections are negligible.
The ambiguity arises as, in principle, the representations, in which
the matter and the gravitational Hamiltonians are quantized, can be
different. However, as we shall discuss in the following, our results
are not sensible on this ambiguity.

We address the genericity of inflation in this
setup, and we investigate whether one can constrain the
parameter space by requiring a sufficiently long inflationary era to
be as likely as possible during the continuum limit of loop quantum
cosmology.

\section{Inflation within loop quantum cosmology}

During the inflationary era the FLRW scale factor $a(t)$ undergoes
an accelerated expansion~\cite{infl2}, ${\rm d}^2a/{\rm
d}t^2>0$. Equivalently, during inflation the universe was dominated by
a fluid with negative pressure which is usually identified as a scalar field.
The scalar field action is
\begin{equation}
\label{action}
S_\phi=\int {\rm d}^4x {\cal L}_\phi=-\frac{1}{2}\int {\rm d}^4x
\sqrt{-g} \left[(\partial \phi)^2-2V(\phi)\right]\ ,
\end{equation}
where the metric is taken to be of the form given in
Eq.~(\ref{metric}).
Inflation is successful in solving the problems which plague the
standard big bang model, provided the slow-roll conditions
\begin{equation}
\left(\frac{\partial V/\partial\phi}{V}\right)^2\ll 1\ \ \mbox{and}\ \
\biggl|\frac{\partial^2 V/\partial\phi^2}{V}\biggr|\ll 1\ ,
\end{equation}
are satisfied for a period of about 60 e-folds, i.e. the final value
of the scale factor, $a_{\rm s}$, must be $a_{\rm s}\approx
\exp(60)a_{\rm i}$, where $a_{\rm i}$ stands for the value of the
scale factor at the beginning of inflation. We define the number of
e-foldings $N$ to be given from $N=\ln(a_{\rm s}/a_{\rm i})$.

\begin{figure}
  \begin{center}
      \includegraphics[scale=0.25]{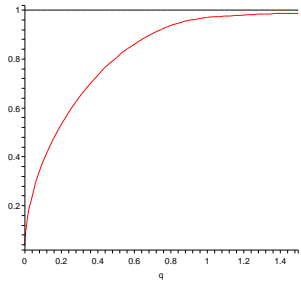}
      \caption{\label{S} $S(q_{\rm G})$
is plotted as a function of $a$, with
 $a_{G}=1$. Notice that $S(q_{\rm G}) < 1$ for all $a$ and
that for $a>>a_{G}$ $S \rightarrow 1$, giving
us the $j = 1/2$ result.}
  \end{center}
\end{figure}

Consider a single-field inflationary model, with an inflaton
field $\phi$ having a potential $V(\phi)$.  The Hamiltonian for $\phi$
obtained from the action $S_\phi$, Eq.~(\ref{action}), reads
\begin{equation}
{\cal H}_\phi=\frac{1}{2}a^{-3}P_\phi^2+a^3V(\phi)\ ,
\end{equation}
where the momentum $P_\phi$ is defined as $P_\phi=-\partial {\cal
L}_\phi/\partial\dot\phi$; an over-dot defines a derivative with
respect to the cosmic time.
We then promote the scalar field Hamiltonian to an operator, thus the
full Hamiltonian for inflation within loop quantum cosmology reads
\begin{equation}
     \hat{\cal{H}}\equiv -3 a{\dot
    a}^2+8\pi G \left[\frac{1}{2} a^{-3}P_\phi^2+ a^3V(\phi) \right]\ .
\end{equation}
As introduced before, we can define the new, classically equivalent,
operator as
\begin{equation}
     \hat{\cal{Q}}\equiv -3 a{\dot
    a}^2+8\pi G \left[ \frac{1}{2}a^{-3(n+1)} a^{3n}P_\phi^2+
     a^{-3m} a^{3(m+1)} V(\phi) \right]\ ,
\end{equation}
where $m,n$ are positive constants.  Upon quantization, $\langle \hat
{\cal Q}\rangle=0$, we obtain,
for $V(\phi)\ll l_{\rm Pl}^{-4}$, in
the slow-roll region, the semi-classical
equation~\cite{martin,Vandersloot2005}
\begin{equation}
\label{hubble}
H^2=\frac{8\pi G S(q_{\rm G})}{3}\left[\frac{1}{2}D_l^{-(n+1)}\dot\phi^2+
D_l^{m}V(\phi)\right]\ ,
\end{equation}
where the Hubble parameter, $H$, is defined as $H \equiv \dot a/a$.
The function $S(q_{\rm G})$ in Eq.~(\ref{hubble}),
\begin{equation}
\label{S(q)}
S(q_{\rm G})={4\over\sqrt q_G}\biggl\{{1\over
  10}[(q_G+1)^{5/2}+\mbox{sgn}(q_G-1) |q_G-1| ^{5/2}]
-{1\over 35}[(q_G+1)^{7/2}- |q_G-1|^{7/2}]\biggr\}~,
\end{equation}
 accounts for the quantization of the gravity part of the Hamiltonian
using a $j_{\rm G} \neq 1/2$ representation~\cite{Vandersloot2005};
the case $j_{\rm G}=1/2$ represent the irreducible representation.  We
have defined as before $q_{\rm G}=a^2/a^2_{\rm G}$.  For $q_{\rm
G}>1$, the function $S(q_{\rm G})$ is $S(q_{\rm G})\approx 1$, while
for small volume, $S(q_{\rm G})~\approx(6/5)\sqrt{q_{\rm G}}$ (see
Figure \ref{S}).  As discussed previously, it is not necessary to use
the same $j_{\rm G}$ representation to quantize both the matter and
gravity parts of the Hamiltonian constraint.  To simplify the
following calculations we set $j_{\rm G}=1/2$ for the quantization of
the gravity part, which implies $S(q_{\rm G})=1$. We discuss the
effects of a more general choice of $S(q_{\rm G})$ in
Section~\ref{subsect:powerlaw}.

Finally, for $V(\Phi)\sim l_{\rm Pl}^{-4}$ and extra term producing a
bouncing in Eq.~(\ref{hubble}) appears~\cite{Ashtekar}. However, our
conclusions apply only if $V(\phi)\ll l_{\rm Pl}^{-4}$.

Using Eq.~(\ref{hubble}) one can write an effective Lagrangian, which leads
to the following
conservation equation for the scalar field $\phi$:
\begin{equation}
\label{cons} \ddot
\phi+\left[3H-(1+n)\frac{\dot D_l}{D_l}\right]\dot
\phi+D_l^{m+n+1}V'(\phi)=0\ ,
\end{equation}
where $V'\equiv\partial V/\partial\phi
$.  We see in
Eqs.~(\ref{hubble}), (\ref{cons}) that there is an ambiguity in
choosing the parameters $m$ and $n$.

\section{The probability to get successful inflation in loop quantum
cosmology}

It has been recently shown~\cite{GT} that it is possible to define a
canonical measure in cosmology. More precisely, it has been
shown~\cite{GT} that the volume of phase space of possible orbits, for
certain inflationary models, is finite if a coarse graining cut-off is
introduced. The authors of Ref.~\cite{GT} argued that two cosmologies
cannot be experimentally distinguished if they differ by a small
amount of spatial curvature, and this removed the divergence present
in the phase space of the relativistic trajectories found in
Ref.~\cite{HawkingPage}.  Considering a finite phase space volume of
possible orbits, they could calculate the fraction of the whole phase
space occupied by inflationary trajectories. In this way they could
define the probability of having inflationary initial conditions in
the framework of classical general relativity.  However, the authors
of Ref.~\cite{GT} allowed the possible trajectories to reach the
Planck scale, where the classical general relativistic Hamiltonian
should not be used.

Here, we modify the proposal of Ref.~\cite{GT}, defining a quantum
gravity cut-off instead of an observational one, namely we calculate
the volume of the phase space of solutions only in the continuum limit
of loop quantum cosmology, {\it i.e.} in the space where
$H^{-1}\gg\sqrt{\gamma}l_{\rm Pl}$ (which we later refer to as the
``quantum gravity cutoff''). We show that the volume is again
finite for the same inflationary models implicitly used in
Ref.~\cite{GT}.

\subsection{Measure: definition}

The canonical cosmological measure of Ref.~\cite{GHS} is given as
follows: As with any phase space we have a symplectic form
\be \label{eq:metric}
\Omega = \sum ^{k}_{i=1} {\rm d}P_{i}\wedge {\rm d} Q^{i}~,
\ee
where $Q_{i}$ and
$P_{i}$ are the dynamical degrees of freedom and their conjugate
momenta, respectively.  The $k^{th}$ power of $\Omega$ gives the
volume element of the space.  The Hamiltonian constraint restricts the
space of trajectories to lie on a $(2k-1)$-dimensional subspace $M$ of
the full phase space, referred to as the multiverse. It can be
shown~\cite{GHS} that $M$ also contains a closed symplectic form
$\omega = \sum _{i=1}^{k-1}{\rm d}P_{i}\wedge{\rm d} Q^{i}$, which is
related to $\Omega$ via,
\be
\Omega = \omega + {\rm d} {\cal H} \wedge {\rm d}t\ \Rightarrow \omega
= \Omega |_{{\cal H} =0}~.
\ee
In particular, this construction can be easily extended to our case by
replacing $\cal H$ with the effective Hamiltonian $\langle \hat{\cal
Q}\rangle$ of the system.

\begin{figure}
  \begin{center}
      \includegraphics[scale=0.25]{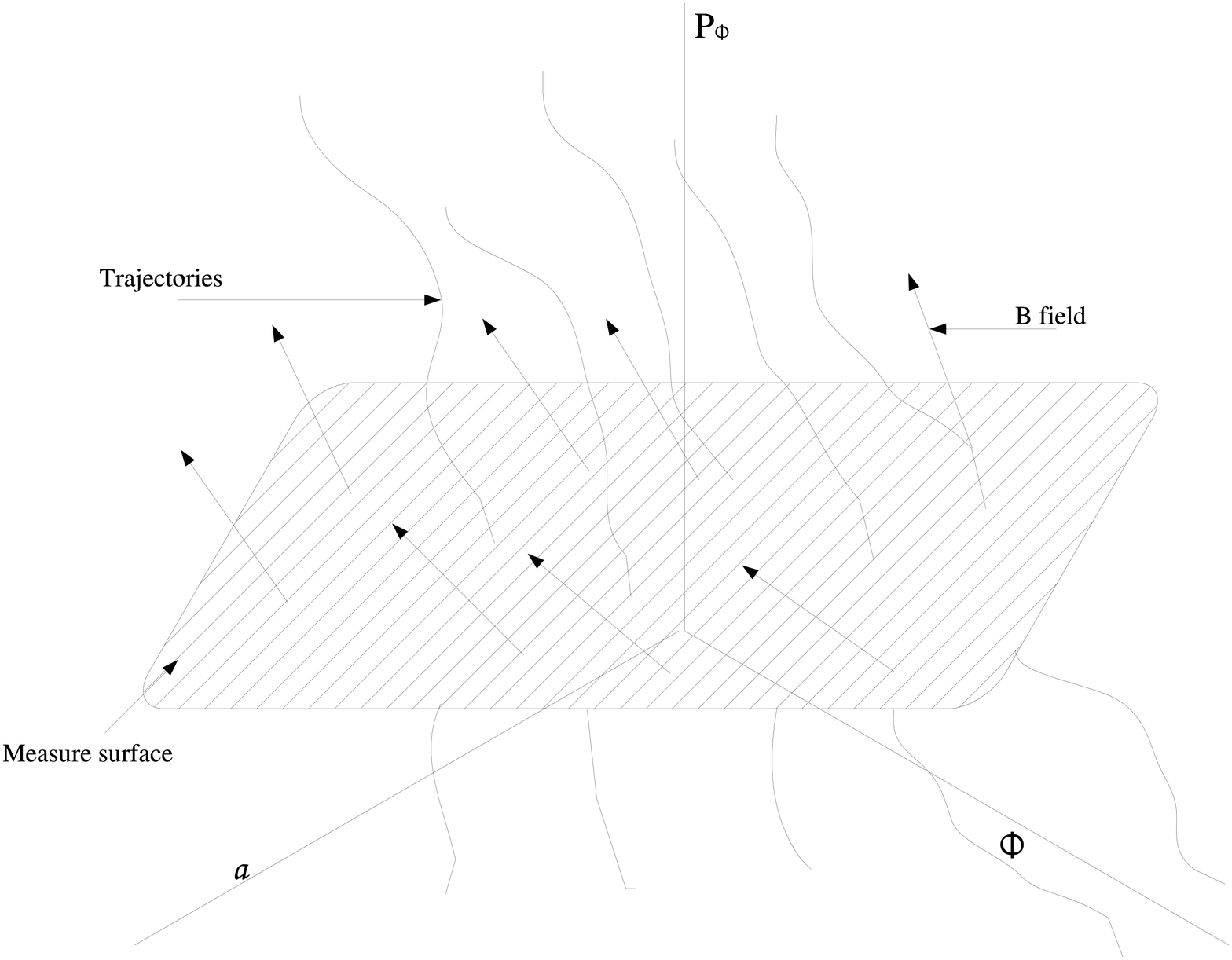}
      \caption{\label{fig:measure} The probability measure is defined
by integrating the ${\bf B}$ field over a constant surface in the
$3$-dimensional phase space produced by using the Hamiltonian
constraint to eliminate one of the dynamical variables.}
  \end{center}
\end{figure}

In FLRW universes containing a scalar field there are only two
canonical variables $(a,\phi)$, so we set $k=2$. Given this symplectic
form it is possible to define a divergenceless field \be B_{a} \equiv
\frac{1}{2} \epsilon_{abc} \omega_{bc}\ ; \ee $\epsilon_{abc}$ is
totally anti-symmetric with $\epsilon_{123}=1$ and $a,b,c = 1, 2,
3$. Each orbit in the phase space is associated with a {\sl line of
force} of $\bf{B}$, i.e.  $\bf{B}$ defines the flow of trajectories
across surfaces in the phase space. We thus define a measure as \be
{\cal N} =\int {\bf B} \cdot {\rm d}{\bf S}~, \ee where $\bf{S}$ is an
open surface where, to ensure that there is no over-counting, the
orbits cross only once. We schematically show this method in
Figure (\ref{fig:measure}).  Since $\bf{B}$ is divergenceless we can
define ${\rm d}\bf{A} = \bf{B}$ and, in the case of a non-disconnected
surface $\bf{S}$, using Stoke's theorem \be {\cal N} = \oint {\bf A}
\cdot {\rm d}{\bf l}~, \ee where ${\bf l}=\partial {\bf S} $ is the
boundary of ${\bf S}$.  The quantity ${\cal N}$ is the canonical
measure of all trajectories crossing topologically equivalent surfaces
which are bounded by $\partial \bf{S}$.

The reader should keep in mind that since Eq.~(\ref{eq:metric})
defines a flat metric on the phase-space, each trajectory has the same
weight, hence we have a measure on the \textit{number} of
trajectories, and not the volume they occupy. This is indeed a crucial
point for the estimation of the probability to have successful
inflation, since such a solution is an attractor, meaning that the
volume occupied by inflationary trajectories decreases, as the
attractor solution is approached. In conclusion, we are confident in
our estimated probability of successful inflation, since we are just
counting the {\it number} of trajectories, and not the phase-space
volume they enclose. 
The measure we use
is certainly not the most general measure of the gravitational
phase-space volume, but just the simplest (a uniform distribution)
one.  The reader should then keep in mind that different distributions
for the phase-space trajectories may be adopted (see for
example~\cite{EKS}).

\subsection{Estimation of the probability}

An estimation of the probability of a set of trajectories ${\cal C}$
in the phase space of all possible trajectories of an Hamiltonian
system, is given by the ratio of the measure of ${\cal C}$ to the
measure of all possible trajectories of trajectories and not the phase
space volume they enclose. For this ratio to be well defined we
require that the trajectories do not cross, as this would lead to a
time dependent measure of ${\cal C}$. The time reversibility of the
system ensures that such crossing does not take place within a finite
time. To define the ratio, we also require the measure of the entire
phase space to be finite. It is well known that in the classical case
this is not true~\cite{HawkingPage}, unless one introduces a coarse
graining cut-off~\cite{GT}. In quantum loop cosmology this classical
divergence is removed since we are restricted to scales above $L_{\rm
Pl}$. However, as we will see in the next section, there is the
possibility of a further divergence associated with the form of the
potential. To ensure that this divergence is not present we must
restrict our attention to a specific class of potentials, a limitation
that is also present in the classical theory of~\cite{GT}.

\subsubsection{The volume of the phase space}
We now turn our attention to the calculation of the total volume
using Eqs.~(\ref{hubble}), (\ref{cons}).
The momentum associated with the scale factor and the scalar field
$\phi$ is
\be
P_{a}=-6a^{2}H ,\
P_{\phi}= a^{3} D_{l}^{-(n+1)} \dot{\phi}~,
\ee
or, in terms of $q=a^{2}/a_{\star}^{2}$ and using
Eq.~(\ref{hubble}),
\be
P_{q} = -  6a_{\star}^{2}q H\ ,\ P_{\phi}=a_{\star}^{3}q^{\frac{3}{2}}
D_{l}^{-(n+1)}   \dot{\phi}   =   D_{l}^{-\frac{n+1}{2}}   a_{\star}^3
q^{\frac{3}{2}} \sqrt{\frac{3H^2}{4\pi G}-2D_{l}^mV}~,
\ee
leading to the constraint
\begin{equation}
H^2>\frac{1}{3}D_{l}^m V~.
\end{equation}
We now calculate the symplectic form
\be
\omega = {\rm d}P_{\phi}\wedge
{\rm d} \phi + {\rm d}P_{q}\wedge {\rm d}q\ ,
\ee
which gives
\begin{eqnarray}
    B_{\phi} &=& -6a_{\star}^{2}q~, \\
    B_{H} &=& \frac{1}{2}a_{\star}^{3} q^{\frac{1}{2}}
\sqrt{\frac{3H^{2}}{4\pi G}-2D_{l}^{m}V} D_{l}^{-\left( \frac{n+1}{2}
\right) } \left[ 3 - \left( n+1 \right) \frac{a_{\rm s}}{D_{l}}
\biggl({\partial D_{l}\over\partial q}\biggr) \right] 
-
\frac{ma_{\star}^{3}q^{\frac{3}{2}}}{\sqrt{\frac{3H^{2}}{4\pi G}
-2D_{l}^{m}V}} D_{l}^{\frac{2m-n-3}{2}} V\biggl({\partial
D_{l}\over\partial q}\biggr)~,  \\ 
    B_{q} &=& -\frac{3Ha_{\star}^{3}
q^{\frac{3}{2}}D_{l}^{-\left( \frac{n+1}{2}
             \right) }}{4\pi G\sqrt{\frac{3H^{2}}{4\pi G}
            -2D_{l}^{m}V}}~.
\end{eqnarray}
An associated vector potential, ${\bf A } = P_{i}\ {\rm d}Q^{i}$,
reads \be {\bf A} = \left[
a_{\star}^{3}q^{\frac{3}{2}}\sqrt{\frac{3H^{2}}{4\pi G}-2D_{l}^{m}V}
D_{l}^{-\left(\frac{n+1}{2} \right)},\ 0,\ -6a_{\star}^{2}qH \right]~.
\ee To calculate the measure, as described in the previous section, we
need to define a surface that is cut only once by the trajectories in
the multiverse. It is convenient to use the surface defined by
$q=q_{\rm s}$, where $q_{\rm s}$ is a constant. We consider expanding
universes, thus from $\langle \hat{\cal Q}\rangle=0$ we get that ${\rm
d} a/{\rm d}t$, and hence ${\rm d} q/{\rm d}t$ is monotonic and
positive if and only if $V>0$, which is guaranteed by the dominant
energy condition\footnote{The dominant energy condition reads $\rho
\geq |p_{i}|$. In the case of a scalar field $\phi$, one has $\rho
\sim {\rm K.E.} + V(\phi)$ and $p_i \sim {\rm K.E.} - V(\phi)$, where
K.E. denotes the kinetic term. Thus, even with the corrections to the
$1/a^{3}$ factor in the kinetic energy terms, the dominant energy condition
ensures that $V(\phi) > 0$.}.  Integrating ${\bf B}$ over a constant
$q=q_{\rm s}$ surface we obtain
\begin{equation}
\label{N}
    \mathcal{N} = - \frac{3}{4\pi G} \int \int
\frac{Ha_{\star}^{3}q^{\frac{3}{2}}_{\rm s} D_{\rm s}^{-\left(
\frac{n+1}{2}\right) } }{\sqrt{\frac{3H^{2}}{4\pi G} - 2D_{\rm s}^{m}
V }} {\rm d}H {\rm d}\phi~,
\end{equation}
where (from now on we drop the $l$ label) we use the notation
$f(a_{\rm s})=f_{\rm s}$ so that $D_{\rm s} \equiv D\left(q_{\rm s}\right)$.

We are now able to integrate Eq.~(\ref{N}).
At this point we introduce the quantum gravity cut-off $H^{-1}\gg
\sqrt\gamma l_{\rm Pl}$. Considering the physical limit \footnote{For
an expanding universe $H>0$.} $H\gg \sqrt{ \frac{8}{3}\pi G D^m V }$
we also obtain in turn that $V\ll l_{\rm Pl}^{-4}$ which avoids, as
anticipated, the quantum gravity bouncing region. We now may perform
the integral over $H$ to obtain
\begin{equation} \label{N2}
    \mathcal{N}=- a_{\star}^{3}q^{\frac{3}{2}}_{\rm s}D_{\rm
    s}^{-\frac{n+1}{2}}  \sum_{k}\int_{\phi_{\rm i}^k}^{\phi_{\rm f}^k}
    \sqrt{ \frac{3}{4\pi\gamma l_{\rm Pl}^{4} }
    -2D_{\rm s}^m V(\phi)}{\rm d}\phi\ ,
\end{equation}
where $[\phi_{\rm i}^k, \phi_{\rm f}^k]$ represent the allowed
(possibly disconnected) ranges for $\phi$ such that ${\cal N}$ is
real.  We note that the integral in Eq.~(\ref{N2}) is by no means
always convergent, as it was assumed in Ref.~\cite{GT}, in the
$D\rightarrow 1$ limit for potentials with only one minima. Its
convergence indeed depends on the choice of $V(\phi)$.  However, a
large class of potentials make it convergent, for example potentials
with only one minima but diverging in the large $\phi$ limit.  These
potentials are phenomenologically very important. Firstly, the
requirement of having a minimum makes it possible to have only one
specific vacuum for the scalar field $\phi$, from which ordinary
matter may be produced. Moreover, in this class of potentials belongs
the case of a massive scalar field with
$V(\phi)=\frac{1}{2}\mu^2\phi^2$, where $\mu$ is a constant mass,
which seems to be the favorite model~\cite{jc} to match the WMAP
data. In Section~\ref{subsect:powerlaw} we will calculate explicitly the
probability to have successful inflation for single-field polynomial
inflation with a potential of the form
$V(\phi)=\frac{\mu^4}{2\alpha!}(\frac{\phi}{\mu})^{2\alpha}$; the
integer constant $\alpha$ is $\alpha\geq 1$ and the self-coupling
constant $\mu$ has dimensions of mass (the $\alpha=1$ case reduces to
the scalar field mass). For the class of potentials for which the
integral in Eq.~(\ref{N2}) converges, the range of allowed $\phi$ is
given by the roots of the equation \be \frac{3}{4\pi \gamma l_{\rm
Pl}^{4} } -2D_{\rm s}^m V(\phi)=0\ , \ee implying
\begin{equation}
    \mathcal{N}=\frac{a_{\star}^{3}q^{\frac{3}{2}}_{\rm s} D_{\rm
        s}^{- (n+1)/2 }}{l_{Pl}^3} \left[\frac{3}{4\pi
        \gamma}\right]^{\frac{\alpha+1}{2\alpha}}\left[\frac{2\alpha!}
{2D_{\rm s}^m}\right]^{\frac{1}{2\alpha}}
        \frac{\sqrt{\pi}}{2}\frac{\Gamma({1\over
        2\alpha})}{\alpha\Gamma(
        \frac{3\alpha+1}{2\alpha})}\left(l_{\rm Pl}\,
        \mu\right)^{\frac{\alpha-2}{\alpha}}\ .
\label{calN}
\end{equation}
This volume is clearly finite as $D$ is a finite function of $q$,
that is bounded below by the cutoff $L_{\rm Pl}$.

\subsubsection{The probability}

We now need to calculate the volume of the phase space that contains
inflationary trajectories. In other words, following exactly the same
calculations as before, we will estimate the measure
\begin{equation} {\cal M}= -\frac{3}{4\pi G} \int \int
    \frac{Ha_{\star}^{3}q^{\frac{3}{2}}_{\rm s} D_{\rm s}^{- (n+1)/2 }
    }{\sqrt{\frac{3H^{2}}{4\pi G}- 2D_{\rm s}^{m} V }} {\rm
    d}H {\rm d}\phi{\Big |}_{\rm inflation}\ .
\end{equation}
Using Stoke's theorem for the path $H=H_{\rm s}={\rm const.}$, we get
\begin{equation} \label{M}
    {\cal M}=- \oint \mid P_\phi\mid {\rm d}\phi{\Big |}_{\rm inflation}\ .
\end{equation}
Note that the above integral is positive as inflation runs from higher
to smaller values of the scalar field $\phi$.
From Eq.~(\ref{hubble}) and using $F^2=H^2D^{-m}$ we obtain
\be \label{HubF}
    F^{2}=8\pi G \left[ \frac{1}{6} D^{-\left(n+m+1 \right)}
    \dot{\phi}^{2} +\frac{1}{3} V \right]\ ,
\ee
or
\be
    \dot{\phi}^{2} = \frac{1}{4\pi G}3D^{n+m+1} F^{2} - 2D^{n+m+1} V\ .
\ee
Using Eq.~(\ref{cons}), after lengthy but
straightforward calculations, we obtain
\be
\label{PhiDot}
\dot{\phi} = - \frac{D^{ n+(m/2)+1 }\left({
    dF\over d\phi}\right) }{4\pi G \left[ 1-(n-m+1)
    \frac{q}{3D} \left({\partial D\over \partial q}\right)
    \right]}~, \ee
for $$1-(n-m+1)
    \frac{q}{3D} \left({\partial D\over \partial q}\right)\neq 0~.$$
Equation (\ref{PhiDot}) implies inflation should happen only in
the $(\usual)>0$ region for an expanding universe ($H>0$) and for
graceful exit from inflation
($\dot \phi<0$). One can easily see this by considering that during slow-roll
$dF/d\phi\sim V'/\sqrt{V}>0$, for expanding cosmologies.
Substituting Eq.~(\ref{PhiDot}) into Eq.~(\ref{M}), we get
\begin{eqnarray}
    {\cal M} &=& - \frac{a_{\rm s}^{3} D_{\rm s}^{ m/2}}{4\pi G
\Big| 1-(n-m+1) \frac{q_{\rm s}}{3D_{\rm s}} \left({\partial
D\over\partial q} |_{\rm s} \right)\Big| } \oint {\rm d} \phi \frac{
{\rm d} F}{ {\rm d} \phi} {\Big |}_{\rm inflation} \nonumber \\ &=& -
\frac{a_{\rm s}^{3} }{4\pi G \Big| 1-(n-m+1) \frac{q_{\rm
s}}{3D_{\rm s}} \left({\partial D\over\partial q}|_{\rm s} \right)
\Big| }\delta H|_{H_{\rm s}, q_{\rm s}}~,
\label{M2}
\end{eqnarray}
where $\delta H|_{H_{\rm s}, q_{\rm s}}$ measures the space of inflationary
trajectories cutting the $H=H_{\rm s}$ surface at $q=q_{\rm s}$.
To calculate $\delta H|_{H_{\rm s}, q_{\rm s}}$, we
substitute Eq.~(\ref{PhiDot}) into Eq.~(\ref{HubF}) and evaluate
on $q=q_{\rm s}$.
Introducing the new positive definite variable
\be \label{eq:A}
    \epsilon = \frac{D^{\frac{n+1}{2}  }}{\Big|1-\left(n-m+1
             \right) \frac{q}{3D} \left({\partial D\over
\partial q}\right)\Big|} F \equiv
            \frac{F }{A}\ ,
\ee
we obtain
\be
    \label{A}
    A_{\rm s}^{2} \epsilon_{\rm s}^{2} = \frac{1}{12 \pi G} \left({
    {\rm d} \epsilon_{\rm s} \over {\rm d} \phi} \right)^{2} + \frac{8
    \pi G}{3} V\ , \ee where $\epsilon_{\rm s}$ and $A_{\rm s}$ are
    $\epsilon$ and $A$ evaluated on $a=a_{\rm s}$.  The slow-roll
    condition in loop quantum cosmology is equivalent to
    $D^{-n-m-1}\dot \phi^2\ll 2 V$, which implies \be
\label{mn}
      \frac{1}{3}\frac{(\partial V/\partial\phi)^2}{V^2}\ll
8\pi G\ A^2.
\ee
Using the variable $\epsilon$ defined above, the slow roll condition
can equivalently be written as $({\rm d}\epsilon_s/{\rm d}\phi)^2\ll
(96\pi G/3) V$.

Let us now consider the perfect slow roll solution ($\epsilon_{\rm sr}$)
such that
\be
A^2_{\rm s}\epsilon^2_{\rm sr}=\frac{8\pi G}{3}V\ .
\ee
The volume of all inflationary trajectories will be an expansion on
small values of $\epsilon$. We will therefore study Eq.~(\ref{A})
using, $\epsilon_{\rm s} \rightarrow \epsilon_{\rm sr} + \delta
\epsilon $, where $\delta \epsilon$ is a small perturbation in
$\epsilon_{\rm sr}$.  This gives
\begin{equation}
    {{\rm d}\delta \epsilon\over {\rm d}\phi}= \frac{3A_{\rm s}^{2}
    \epsilon_{\rm s} \delta \epsilon}{ \sqrt{ \frac{ 3A_{\rm s}^{2}
    \epsilon_{\rm s}^{2}}{ 4\pi G} -2V}}\ .
\end{equation}
If we now define the function
\be
N_{\rm s}=\int^{a_{\rm s}}_a \frac{{\rm d}\tilde{a}}{\tilde{a}}\ ,
\ee
we have
\be
    {{\rm d} N_{\rm s}\over {\rm d}\phi} = \frac{H_{\rm s}}{\dot{\phi_{\rm
    s}}}= \frac{\epsilon_{\rm s} A_{\rm s} D_{\rm s}^{-\left(
    \frac{n+1}{2} \right) } }{ \sqrt{ \frac{3A_{\rm
    s}^{2}\epsilon_{\rm s}^{2}}{ 4\pi G} -2V}}\ ,
\ee so that
\be
{{\rm d} \delta \epsilon\over {\rm d} N_{\rm s}} = 3 \Big| 1
- \left( n -m +1 \right) \frac{q_{\rm s}}{3D_{\rm s}} \left({\partial
D\over\partial q}{\Big |}_{\rm s}  \right) \Big| \delta \epsilon\ .
\ee
We finally have
\be
\delta \epsilon=C \exp \left( 3\Big| 1 - \left( n
-m +1 \right) \frac{q_{\rm s}}{3D_{\rm s}} \left({\partial
D\over\partial q}{\Big |}_{\rm s}  \right) \Big| N_{\rm s} \right) \ ,
\ee
where $C$ is a constant. Taking $a_{\rm s}$ to be the scale at the end of
inflation and measuring $N_{\rm s}$ from $a_{\rm s}$ to the beginning of
inflation, we get
\be
\label{delta1} \delta \epsilon_{\rm i}=\delta
\epsilon_{\rm f} \exp \left( 3\Big| 1 - \left( n -m +1 \right)
\frac{q_{\rm s}}{3D_{\rm s}} \left({\partial D\over\partial
q}{\Big |}_{\rm s}  \right) \Big| N \right) \ , \ee where $\delta
\epsilon_{\rm f}$ and $\delta \epsilon_{\rm i}$ are the perturbation
evaluated at the end and beginning of inflation, respectively, and $N$
stands here for the total number of e-foldings during the slow-roll.

By iterating Eq.~(\ref{A}) it is easy to see that
\be \label{eq:iter}
\epsilon_{\rm s} \approx \frac{1}{A_{\rm s}}\sqrt{\frac{8\pi G
V(\phi)}{3} } \left[ 1+ \frac{1}{96A_{\rm s}^{2}\pi G} \left(
\frac{1}{V(\phi)}\ \frac{ \partial V(\phi) }{\partial \phi}
\right)^{2} +\dots \right]~.
\label{expe}
\ee
Using this expansion we can write,
\be \label{de}
 \delta \epsilon_{\rm i} \approx \frac{1}{12 A^{3}_{\rm s}}
 \sqrt{\frac{V(\phi_{\rm i})}{24\pi G}}
 \left( \frac{1}{V(\phi_{\rm i})}\ \frac{ \partial V(\phi)
 }{\partial \phi} \Big|_{\rm i} \right)^{2}~.
\ee
From Eq.~(\ref{delta1}) we have
\be
\delta \epsilon_{\rm f}=\delta \epsilon_{\rm i} \exp \left(- 3\Big| 1 -
\left( n -m +1 \right) \frac{q_{\rm s}}{3D_{\rm s}} \left({\partial
D\over\partial q}{\Big |}_{\rm s} \right) \Big| N \right) \ .
\ee
Since we
perturbed $\epsilon$ on the constant $a=a_{\rm s}$ surface, we have
that
\be
\delta \epsilon_{\rm f} = \frac{D_{\rm s}^{
\frac{n-m+1}{2} } }{\Big| 1- (n-m+1)\frac{q_{\rm s}}{3D_{\rm s}}
\left({\partial D\over\partial q}|_{\rm s} \right) \Big| } \delta
H|_{H_{\rm s},a_{\rm s}}~.
\ee
Putting these together and substituting into Eq.~(\ref{M2}) gives the
measure of the trajectories that inflate,
\begin{equation}
    {\cal M} = \frac{a_{\star} ^{3}q^{\frac{3}{2}}_{\rm s} }{4\pi G}
D_{\rm s} ^{-\left( \frac{n-m+1}{2} \right) } \delta \epsilon_{\rm i}
 \exp \left( -3 \Big| 1 - (n-m+1)\frac{q_{\rm
s}}{3D_{\rm s}} \left( \frac{\partial D}{\partial q} \Big|_{\rm s}
\right) \Big| N \right)~.
\end{equation}
Thus, the probability of getting $N$ or more e-foldings is
\begin{equation}
\label{probability}
{\cal P}(N)= \frac{1}{\cal N} \frac{
a_{\star}^{3}q^{\frac{3}{2}}_{\rm s} D_{\rm s}^{-\frac{n-m+1}{2} }
}{4\pi G} \delta \epsilon_{\rm i} \nonumber \label{exp} 
\exp\left( -3 \Big| 1 - \left( n -m +1 \right) \frac{q_{\rm s}}{3D}
\left({\partial D \over\partial q}\Big|_{\rm s} \right) \Big| N\right) \ .
\end{equation}

\subsubsection{Probability for polynomial potentials}
\label{subsect:powerlaw}
As we already discussed, our measure is valid only for a sub-class of
possible inflationary potentials. In particular, we now discuss
polynomial potentials
\be
\label{powerlaw}
V(\phi)=\frac{\mu^4}{2\alpha!}\ \left(\frac{\phi}{\mu}\right)^{2\alpha}~.
\ee
From the slow-roll conditions and Eq.~(\ref{eq:iter}) we obtain
\be
\frac{H}{\dot \phi}\approx -8\pi G \left( \frac{1}{V(\phi)}
 \frac{\partial V(\phi)}{\partial \phi} \right)^{-1}
 \frac{[\usual]}{D^{n+1}}\ .
 \ee
Integrating the identity $(1/a)=(H/\dot
\phi)({\rm d}\phi/{\rm d}a)$, we obtain that for polynomial potentials
\be
\label{int-pl}
-\int^{a_{\rm s}}_{a_{\rm i}}\frac{D^{n+1}}{[\usual]}
\frac{{\rm d}a}{a}\approx
\frac{4\pi G}{\alpha}(\phi^2_{\rm f}-\phi^2_{\rm i})\ ,
\ee
where, as before, we took $a_{\rm s}$ as the scale factor at the end of
inflation and $a_{\rm i}$ the scale factor at the begin of inflation.
Equation (\ref{int-pl}) implies that slow-roll (for which
$[\usual]>0$) ends when $\phi_{\rm f}<\phi_{\rm i}$.  In particular,
using the standard definition for the end of inflation,
$\phi_{\rm f}\ll \phi_{\rm i}$, we obtain the following good approximation
\be
\phi_{\rm i}\approx \left(\frac{\alpha}{4\pi
G}\int^{a_{\rm s}}_{a_{\rm i}}\frac{D^{n+1}}
{[\usual]}\frac{{\rm d}a}{a}\right)^{1/2}\ .
\ee
Note that for $\alpha=1$ and $D=1$ we recover the usual result
$\phi_{\rm i}\approx \sqrt{N/(4\pi G)}$.

We are now able to evaluate $\delta\epsilon_{\rm i}$. Using
Eq.~(\ref{de}) we get
\be
\label{deltaepsiloni}
\delta\epsilon_{\rm i}\approx \frac{l_{\rm
Pl}^{-1}}{\sqrt{2\alpha!}}\frac{\alpha^2}{3 A_{\rm
s}^3\sqrt{24\pi}}\left(\frac{\mu}{\phi_{\rm i}}\right)^{2-\alpha}\ .
\ee
Finally, from Eqs.~(\ref{calN}), (\ref{probability}) and
(\ref{deltaepsiloni}), the probability ${\cal P}(N)$
of having  $N$ e-folds of slow-roll inflation reads
\begin{equation}
\label{explicitprob}
{\cal P}(N)\approx \beta^2 \left(\frac{\mu}{\phi_{\rm
i}}\right)^{2-\alpha}\left(l_{\rm
Pl}\mu\right)^{\frac{2-\alpha}{\alpha}}
\exp \left(-3 \Big| 1 -
\left( n -m +1 \right) \frac{q_{\rm s}}{3D} \left({\partial D
\over\partial q}\Big|_{\rm s} \right) \Big| N \right)~,
\end{equation}
where
\be\label{prob}
 \beta^2=\frac{\alpha^3}{144}
\left[\frac{2}{3\pi(2\alpha!)}\right]^{\frac{\alpha+1}{2\alpha}}
2^{\frac{\alpha+2}{2\alpha}}\pi^{\frac{\alpha-1}{\alpha}}
\gamma^{\frac{\alpha+1}{2\alpha}}
\frac{\Gamma(\frac{3\alpha+1}{2\alpha})}{\Gamma({1\over 2\alpha})}
D_s^{m\frac{\alpha+1}{2\alpha}} A_s^{-3}~.
\ee
The above probability
changes qualitatively for renormalizable ($\alpha\leq 2$) and
non-renormalizable ($\alpha>2$) potentials. We will concentrate
on  renormalizable potentials~\cite{pea}, thus
$\alpha=1,2$.

The above calculation can be repeated using $S(q_{\rm G}) \neq 1$ to
give
\begin{equation}
%{eqnarray}
{\cal P}(N)\approx \beta^2 \left(\frac{\mu}{\phi_{\rm
i}}\right)^{2-\alpha}\left(l_{\rm
Pl}\mu\right)^{\frac{2-\alpha}{\alpha}} \biggl[S\biggl({a^2\over
a_G^2}\biggr)\biggr]^{\frac{\alpha+4}{4\alpha}} 
%\nonumber\\
%&&
\times \exp \left(-3 \Big| 1 - \left( n -m +1 \right) \frac{q_{\rm
s}}{3D} \left({\partial D \over\partial q}\Big|_{\rm s} \right) \Big|
N \right)~,
\end{equation}
%{eqnarray}
where now, \be \phi_{\rm i}\approx \left(\frac{\alpha}{4\pi
G}\int^{a_{\rm s}}_{a_{\rm i}}\frac{D^{n+1}} {S\biggl({a^2\over
a_G^2}\biggr)\biggl[\usual\biggr]}\frac{{\rm d}a}{a}\right)^{1/2}\ .
\ee 
Since $S(q_{\rm G})<1$ it is clear that choosing $j_{\rm G} \neq 1/2$
slightly reduces the probability of inflation.
However, for $a_{\rm i}>a_{{\rm G}}$, $S(q_{\rm G})$ is, well approximated
by 1, thus the conclusions for the $j_{\rm
G}=1/2$ case remain qualitatively unchanged. We shall therefore
consider in the following only the $j_{\rm G}=1/2$ fundamental
representation case.

\subsection{Estimation of the probability}

Let us first note that since we have assumed conditions favoring the
onset of inflation (i.e. FLRW universes), finding a high probability
in this context only gives a necessary condition for inflation to be
likely. In order to have enough inflation we require
\be
\label{nec1}
e^{60}\approx \frac{a_{\rm s}}{a_{\rm i}}< \frac{a_{\rm s}}{a_{\rm
Pl}}\ ; \ee $a_{\rm Pl}$ is the minimal scale which can be probed in
our approach.  There are the following possibilities, either $a_{\rm
s}\leq a_\star$, or $a_{\rm s}>a_\star$.  In loop quantum gravity $j$
gives the scale for which semiclassical effects can be
observed. Obviously $j$ cannot be too large otherwise we will probe
quantum gravity at everyday scales. In fact, particle physics
experiments restrict $j<10^{20}$~\cite{martin}. With this bound for
$j$ it is easy to show that the only possibility for which the
necessary condition, Ed.~(\ref{nec1}), is satisfied is in the large
$a_{\rm s}$ limit, {\em i.e.} $a_{\rm s}\gg a_\star$.  With this
inequality we can expand any function evaluated at $a_{\rm s}$ in the
large $q$ limit.

\begin{figure}
  \begin{center}
      \includegraphics[scale=0.25]{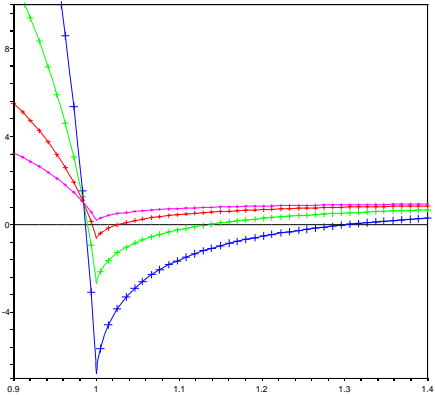}
    \caption{\label{fig1} The function $\left[1-(n-m+1)\frac{q
 }{3D} \left({\partial D\over \partial q}\right)\right] $
is plotted as a function of $a$, with
 $a_{\star}=1$ and $l=0.75$, for $n-m+1=-1,-4,-9,-19$ (small to large crosses
respectively). This is the term that determines the suppression of the
probability of having $N$ e-foldings of inflation. Notice that for $n-m+1
< -2$ we have two zeros, although this number is weakly dependent on $l$.}
  \end{center}
\end{figure}

\begin{figure}
  \begin{center}
      \includegraphics[scale=0.25]{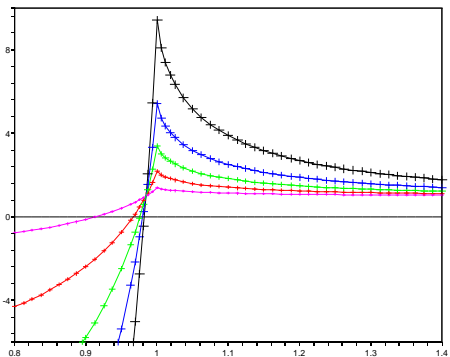}
      \caption{\label{fig2} The function $\left[1-(n-m+1)\frac{q
 }{3D} \left({\partial D\over \partial q}\right)\right]$
is plotted as a function of $a$, with
 $a_{\star}=1$ and $l=0.75$, for $n-m+1=1,3,6,11,21$ (small to large cross
respectively). This is the term that determines the suppression of the
probability of having $N$ e-foldings of inflation. Notice that for $n-m+1 >0$
we have only one zero that is always close to $a_{\star}$ }
  \end{center}s
\end{figure}

\begin{figure}
  \begin{center}
      \includegraphics[scale=0.25]{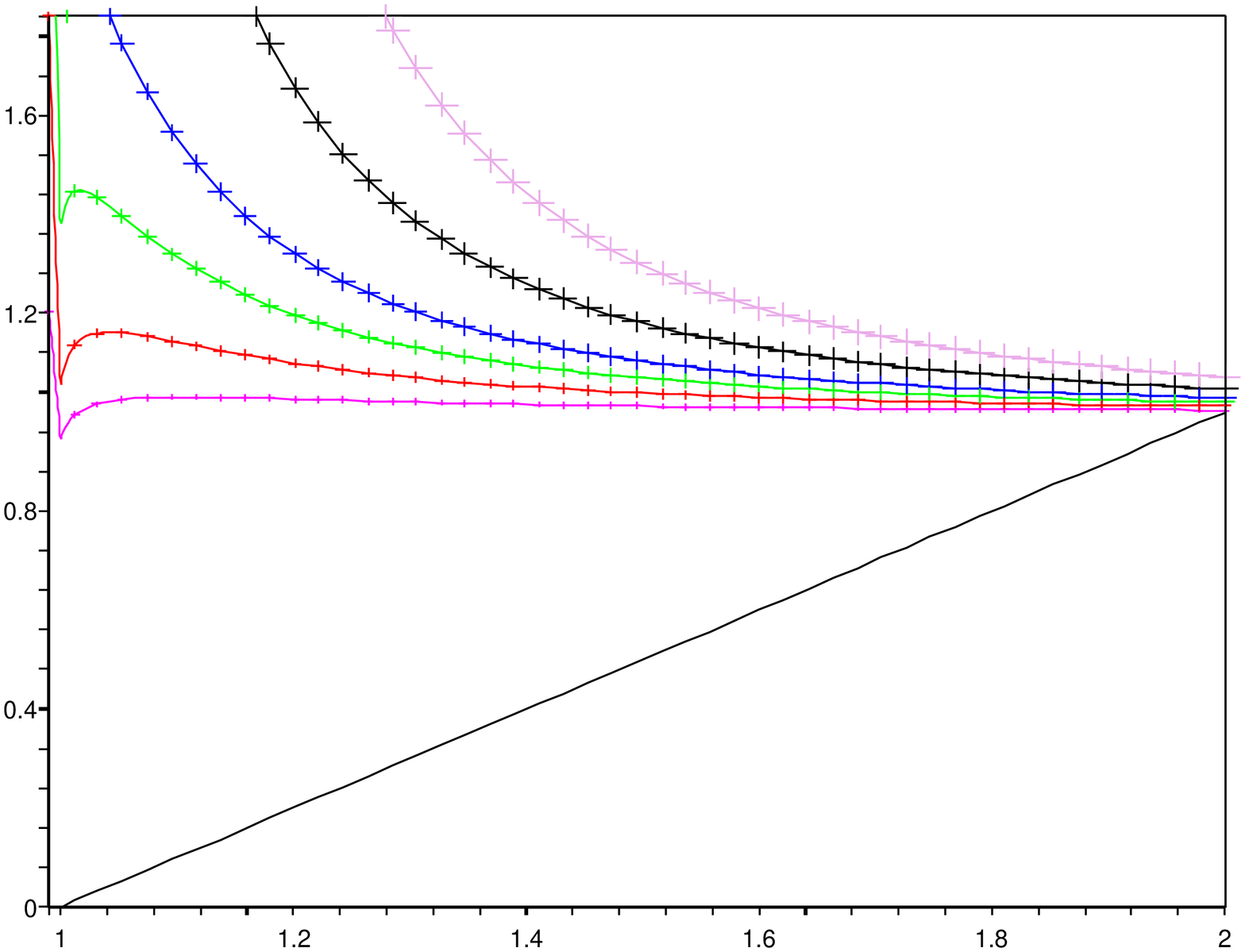}
      \caption{\label{fig4} The function $D^{n+1} \left[
 1-(n-m+1)\frac{q }{3D} \left({\partial D\over \partial q}\right)
 \right]^{-1}$ is plotted as a function of $a$, with $a_{\star}=1$ and
 $l=0.75$, for $n-m+1=1,3,5,7,11,16$ (small to large crosses
 respectively).  Also plotted is an example of $(a-a_{\rm
 i})/(a_s-a_i)$ for $a_i=a_{\star}=1$ and $a_s=2$, which allows us to
 calculate a lower bound on $\phi_{\rm i}$, which appears as one of
 the coefficients of the probability. }
  \end{center}
\end{figure}

Let us discuss the magnitude of the probability, Eq.~(\ref{explicitprob}),
for renormalizable potentials:
\begin{enumerate}
\item The probability, Eq.~(\ref{explicitprob}), is suppressed by a factor
$$\exp\left(-3\left[\usual\right]\Big|_{a_{\rm s}} N\right)~.$$ In
order to make the probability high enough, one could naively think
that by just finding appropriate values for $n, m, a_{\rm s}$ which
make the exponent of ${\cal O}(1)$, one can overcome the negative
result of Ref.~\cite{GT}. However,
$$\beta^2\propto \left[\left(\usual\right)\Big|_{a_{\rm s}}\right]^{3}$$
acts against this reduction. In fact, increasing the value of the
exponential would actually make the probability of having a successful
slow-roll inflation closer to zero. Therefore, the higher estimation of
the probability may be found only when
$$\left(\usual\right)\Big|_{a_{\rm s}}\sim {\cal O}(1)~.$$

\item The second term to take care of, is the factor $l_{\rm Pl}\mu$.
However, natural conditions for inflation \cite{pea} require the
scalar field mass to be much lower than the Planck mass, {\it i.e.}
$l_{\rm Pl}\mu\ll 1$.  For $\alpha=1$ the probability is therefore
suppressed by the factor $(l_{\rm Pl}\mu)^2$.

\item The most interesting term is the factor $\mu/\phi_{\rm i}$. We
have already discussed that $|\usual|$ has to be far from zero at the
end of inflation. However, in principle it can be close to zero at the
beginning of, or during inflation, in compatibility with
Eq.~(\ref{mn}). In this case the probability is again suppressed as
the integral defining $\phi_{\rm i}$ contain
$\left|\usual\right|^{-1}$.
In order to increment the probability  $a_{\rm i}$ has to be then far
from the zero of $|\usual|$. To have $\phi_{\rm i}$ as small as possible
we therefore need to have $D$, in the range $[a_{\rm i},a_{\rm s}]$ as
small as possible. In particular, in order to improve the classical
result $\phi_{\rm i}\sim \sqrt{N}$, one needs
$$D^{n+1} \left|\usual\right|^{-1} < 1.$$ In the case of
$n-m+1\lesssim -\epsilon_-$ where, numerically, one can show that
$-1.5<\epsilon_-<-4$ dependently on the value of $l$ taken, we have
two zeros of $|\usual|$ (see Figure (\ref{fig1})), $a_{{\rm
c}_1}\approx a_{\star}$ and $a_{{\rm c}_2}>a_{\star}$, so $a_{\star} <
a_{\rm i} < a_{\rm s}$.  In this region $D>1$ (see Figure
(\ref{fig3})) and
$$|\usual| < 1~,$$ hence $\phi_{\rm i} > \sqrt{N/4\pi G}$.

For the case in which $n-m+1 > \epsilon_+$, where, numerically, one
can show that $0<\epsilon_+<1$ dependently on the value of $l$ taken,
we have only one zero of $|\usual|$ that is always close to
$a_{\star}$ (see Figure (\ref{fig2})).  In this case the function
$D^{n+1} \left| \usual \right| ^{-1} $ can be less or greater than
one, depending on the choice of $n$ and $a$, however in the region we
are concerned with it is always greater than $(a-a_{\rm i}) / (a_{\rm
s} - a_{\rm i} )$ (see Fig.~(\ref{fig4})).

This gives, \be \phi_{\rm i} > \left[ \frac{\alpha}{4 \pi G} \left(
1-\frac{a_{\rm i} N}{a_{\rm s} - a_{\rm i}} \right) \right]^{1/2}
\approx \left[ \frac{\alpha}{4 \pi G} \right]^{1/2}\ , \ee where we
use the fact that $a_{\rm s} \gg a_{\rm i}$.

In the case $\epsilon_-<n-m+1<0$ we do not have any zeros of
$|\usual|$, and $a_{\rm i}$ is only restricted to be above the Planck
scale.  In this case we can consider the following estimation:
\begin{eqnarray}
\label{est1}
\phi_{\rm i}&\approx& \left(\frac{\alpha}{4\pi
G}\int^{a_{\rm s}}_{a_{\rm i}}\frac{D^{n+1}}
{|\usual|}\frac{{\rm d}a}{a}\right)^{1/2}\cr
&>& \left(\frac{\alpha}{4\pi
G}\int^{a_{\rm s}}_{a_{\star}}\frac{D^{n+1}}
{|\usual|}\frac{{\rm d}a}{a}\right)^{1/2}\ .
\end{eqnarray}
By the arguments above we have \be \phi_{\rm i} > \left(
\frac{\alpha}{4 \pi G} \int^{a_{\rm s}}_{a_{\star}} \frac{{\rm d}a}{a}
\right) ^{1/2}\ .  \ee But as described above, particle physics
experiments restrict $a_{\star}< 10^{10}\sqrt\gamma /\sqrt{3}$, which
implies
\begin{eqnarray}
\phi_{\rm i} &>& \left[ \frac{\alpha}{4 \pi G} \left( N - \ln  \left(
10^{10}/\sqrt{3}\right) \right) \right]^{1/2}
\nonumber \\
&>& \left[ \frac{\alpha}{4\pi G} \left( N - 22.5 \right)\right] ^{1/2}\ .
\end{eqnarray}

Finally, in the case $0<n-m+1<\epsilon_+$ we still don't have any zeros,
however, the function $\frac{D^{n+1}} {|\usual|}$ approaches $1$ from above,
so
\be
\phi_{\rm i}>\sqrt{\frac{N}{4\pi G}}\ .
\ee

We have therefore shown that the probability of having slow-roll
 inflation is not significantly improved by the factor $\mu/\phi_{\rm i}$,
for values of $N> 22.5$.

It has been shown~\cite{mb}, that quantum loop cosmology can lead to
a period of {\it super-inflation} during which the scalar field is driven
up its potential. Since this period does not satisfy the slow-roll conditions
it is not accounted for by our analysis. However, perturbation theory is
unstable in this {\it super-inflationary} epoch~\cite{Tsujikawa:2003vr} and
hence, to produce the observed CMB anisotropies we still
require approximately $60$ e-foldings of standard slow-roll inflation.

\item In the probability, Eq.~(\ref{explicitprob}), the factor
 $\beta^2\propto D_{s}^{\frac{m(\alpha+1)-\alpha(n+1)}{2\alpha}}$ (see
 the definition of $\beta$ in (\ref{prob}) combined with the
 definition of $A$ in (\ref{eq:A})), can in principle be big for large
 values of $m$ such that $m>\frac{\alpha}{\alpha+1}(n+1)$. This is due
 to the fact that the function $D$ approaches $1$ from above. We can
 estimate the magnitude of this factor by expanding $D$ for large
 $q$'s. We have (note that $l<1$) \be D_s \simeq 1+ 3\frac{2-l}{20}
 q_s^{-2}\ .  \ee We have already discussed that, in all cases,
 $a_{\rm i}>\sqrt\gamma$ so, for $N=60$, we have $q_{\rm s} >
 10^{32}$.  In this case, if we want that the factor
 $D_{s}^{\frac{m(\alpha+1)-\alpha(n+1)} {2\alpha}}$ to at least
 overcome the exponential suppression $\exp (-180)$ we have the
 necessary condition \be
 \frac{m(\alpha+1)-\alpha(n+1)}{2\alpha}\gtrapprox 10^{110}\ , \ee
 which is possible only for very `un-natural' parameters of the loop
 quantum cosmology~\cite{martinetal}.
\end{enumerate}

\section{Conclusion and discussion}

Cosmological inflation remains still the most appealing candidate in
solving the puzzles of the standard hot big bang model. However,
inflation must prove itself generic. This is an old question which has
been already faced in the past~\cite{piran, calzettamairi}.  Recently,
this issue came back with the study of Ref.~\cite{GT}, where it was
concluded that successful inflation is unlikely. More precisely, it
was argued that the probability to have successful inflation is
exponentially suppressed by the number of e-folds. Clearly, since such
a conclusion leads necessarily to very severe implications, one should
be sure for its validity and generality.  The study of Ref.~\cite{GT}
has, to our opinion, a {\sl weak} point. Classical physics has been
applied all the way to very early times, and therefore very small
scales as compared to the Planck length, a regime where quantum
corrections can no longer be neglected. The estimation of the
probability to have successful inflation should, to our opinion, be
done taking quantum corrections into account. This is indeed the aim
of our work. More precisely, in this paper we have addressed the
question of how likely is the onset of inflation during the continuum
phase of loop quantum cosmology.

Modifying the canonical measure introduced in Ref.~\cite{GHS}, so that
it is applicable in the context of loop quantum cosmology, we have
found that it is not probable to get sufficiently long single-field
inflation, for the phenomenologically favorite inflationary models,
unless we accept extreme values for the ambiguity parameters $m, n$.
Since, during the semi-classical era of loop quantum cosmology, the
filed $\phi$ can depart from the minimum of its
potential~\cite{martinetal}, one may think that this could improve the
classical probability of inflation~\cite{Singh_Vandersloot_etal}. This
however is not what we obtain from our analysis, and the reason is
that the same mechanism which forces $\phi$ away from its minimum, it
will also increase $\dot{\phi}$, which would tend to reduce the
probability of the onset of slow-roll.  In conclusion, our results
show that overall, quantum loop cosmology does not significantly
improve on the classical probability, unless one accepts extreme
values of the ambiguity parameters.

Our results hold for single-field inflation with potentials which
makes the volume of the phase space of possible trajectories
finite. For example, inflationary models with potentials of the form
$V(\phi)\sim \phi^{2\alpha}$ (with $\alpha$ an integer number) are
within the class of models we have studied here.  Our result implies
limitations in the form of inflationary models within loop quantum
cosmology. Since eventually the form of the inflationary model will be
dictated from a fundamental theory, this {\sl freedom} in modeling the
inflationary potential will be alleviated.

From the analysis presented here, it is clear that for the classes of
models studied, the values $m=n=0$ do not lead to a successful
inflationary model. This implies important consequences.  In the
literature on loop quantum cosmology, the ambiguity parameters $m, n$
have been usually both set equal to zero. Clearly, in this context
successful cosmological inflation cannot take place in the
semi-classical regime. Actually, one expects to constrain the ambiguity
parameters by investigating the observational consequences to which
inverse volume operators lead.  For example, to study cosmological
perturbations in loop quantum cosmology, one should perturb both, the
gravitational as well as the matter parts, about the homogeneous
background. This has been only recently accomplished. In
Ref.~\cite{mb} inhomogeneous cosmological perturbation equations have
been derived without neglecting corrections in the gravitational part,
thus treating both gravitational and matter terms on equal
footing. This is indeed the appropriate framework to study
cosmological perturbations in the context of loop quantum
cosmology~\cite{Dittrich_etal}. However, also in this study,  $m, n$
have been set equal to zero, which as we have shown here do not lead
to successful inflation.

We have also analyzed the probability to have successful
inflation for arbitrary values of the ambiguity parameters $m, n$. Our
study has shown that successful inflation can be realized only for
extreme values of the parameters; a result which goes against the {\sl
spirit} of inflation.

Our findings do not imply that inflation itself is improbable. What we
have shown here is that, at least in the case of the semi-classical
regime of loop quantum cosmology and therefore of general relativity,
inflation is not as general as it is usually assumed. Thus, one has to
address inflation in full quantum gravity, or in a string theory
context.

\section*{Acknowledgments}
We are grateful to Martin Bojowald for a number of valuable
discussions on loop quantum cosmology. We thank Nick Mavromatos for
comments on Ref.~\cite{GT}. C.~G.  thanks the Department of Physics at
King's College London for hospitality during part of this work. The
research of W.\ N.\ and M.\ S. was supported in part by the European
Union through the Marie Curie Research and Training Network
UniverseNet (MRTN-CN-2006-035863).


\section*{References}
\begin{thebibliography}{10}
\bibitem{infl1}
A.\ Guth, Phys.\ Rev.\ D {\bf 23}, 347 (1981).

\bibitem{stringgas} R.\ H.\ Brandenberger and N.\ Shuhmaher, JHEP {\bf
0601}, 074 (2006).

\bibitem{cyclic} P.~J.~Steinhardt and N.~Turok, Science {\bf 296}
(2002) 1436.

\bibitem{sling} C.~Germani, N.~E.~Grandi and A.~Kehagias, {\sl A stringy
alternative to inflation: The cosmological slingshot scenario},
[arXiv:hep-th/0611246].

\bibitem{piran}
T.\ Piran, Phys.\ Lett. B {\bf 181}, 238 (1986); D.\ S.\ Goldwirth,
Phys.\ Rev.\ D {\bf 43}, 3204 (1991).

\bibitem{calzettamairi}
E.\ Calzetta and M.\ Sakellariadou, Phys.\ Rev.\ D {\bf 45}, 2802 (1992);
{\sl idem}  Phys.\ Rev.\ D {\bf 47}, 3184 (1993).

\bibitem{GT} G.\ W.\ Gibbons and N.\ Turok, {\sl The measure problem
in cosmology}, [hep-th/0609095].

\bibitem{rovelli} C.\ Rovelli, {\sl Quantum Gravity}, Cambridge
University Press, Cambridge UK, 2004.

\bibitem{tie}
T.\ Thimann, Lect.\ Notes Phys.\ {\bf 631}, 41 (2003).

\bibitem{martin} M.\ Bojowald, {\sl Loop quantum cosmology},
[gr-qc/0601085].

\bibitem{martinetal}
M.\ Bojowald, J.\ Lidsey, D.\ J.\ Mulryne, P.\
Singh and R.\ Tavakol, Phys.\ Rev.\ D {\bf 70}, 043530 (2004).

\bibitem{aps06}
A.\ Ashtekar, T.\ Pawlowski and P.\ Singh,
Phys.\ Rev.\ D {\bf 74}, 084003 (2006).

\bibitem{mb06}
M.\ Bojowald,  Gen.\ Rel.\ Grav.\ {\bf 38}  1771 (2006).

\bibitem{Vandersloot} K.\ Vandersloot, Phys.\ Rev.\ D {\bf 75} 023523
(2007)

\bibitem{ABL} A.\ Ashtekar, M.\ Bojowald, J.\ Lewandowski, Adv.\
Theor.\ Math.\ Phys.\ {\bf 7} 233 (2003)

\bibitem{Tsujikawa:2003vr} S.~Tsujikawa, P.~Singh and R.~Maartens,
Class.\ Quant.\ Grav.\ {\bf 21} (2004) 5767.

\bibitem{infl2}
A.\ Linde, Phys.\ Lett.\ B{\sl 108}, 389 (1982).

\bibitem{Vandersloot2005}
K.\ Vandersloot, Phys.\ Rev.\ D {\bf 71} 103506 (2005)

\bibitem{Ashtekar} A.~Ashtekar, T.\ Pawlowski, P.\ Singh,
and K.\ Vandersloot,  Phys.\ Rev.\ D {\bf 75}, 024035 (2007).

\bibitem{EKS}
W.\ R.\ Stoeger, G.\ F.\ R.\ Ellis and U.\ Kirchner
Mon.\ Not.\ Roy.\ Astron.\ Soc.\ {\bf  347} (2004) 921.

\bibitem{HawkingPage}
S.\ W.\ Hawking, D.\ N.\ Page, Nucl.\ Phys.\ B {\bf 298}, 789 (1988)

\bibitem{GHS} G.\ W.\ Gibbons, S.\ W.\ Hawking, J.\ M.\ Stewart,
Nucl. Phys. B {\bf 281}, 736 (1986)

\bibitem{jc}
J.\ Martin and C.\ Ringeval, JCAP {\bf 0608}, 009 (2006).

\bibitem{Singh_Vandersloot_etal} P.\ Singh, K.\ Vandersloot, G.\ V.\
Vereshchagin, Phys.\ Rev.\ D {\bf 74}, 043510 (2006)

\bibitem{pea} J. A. Peacock, {\sl Cosmological Physics},
[Cambridge University Press, Cambridge (2002)].

\bibitem{mb} M.\ Bojowald, H.\ H.\ Hernández, M.\ Kagan, P.\ Singh,
A.\ Skirzewski, {\sl Hamiltonian cosmological perturbation theory with
loop quantum gravity corrections}, [gr-qc/0609057].

\bibitem{Dittrich_etal} B.\ Dittrich, J.\ Tambornino, {\sl Gauge
invariant perturbations around symmetry reduced sectors of general
relativity: applications to cosmology}, [gr-qc/0702093].

\end{thebibliography}
\end{document}